\documentclass[twocolumn,secnumarabic,amssymb, nobibnotes, aps, prc, superscriptaddress, nobalancelastpage,longbibliography]{revtex4-2}

 \usepackage{graphicx}
\usepackage{bm}
\usepackage{amsmath} 
\usepackage{braket} 
\usepackage{epsfig}
\usepackage{tensor}
\usepackage{CJKutf8}
\usepackage[version=4]{mhchem}
\usepackage{titlesec}
\usepackage{ifthen}
\usepackage{sidecap}
\usepackage{listings}
\usepackage{array}
\usepackage{floatrow}
\usepackage{multirow}
\usepackage{enumitem}
\usepackage[para,online,flushleft]{threeparttablex}
\usepackage{float}
\floatstyle{plaintop}
\restylefloat{table}
\usepackage{booktabs}

\usepackage{hyperref}
\hypersetup{breaklinks=true,colorlinks=true,linkcolor=blue,citecolor=blue,filecolor=magenta,urlcolor=cyan}

\usepackage[all]{hypcap}

\usepackage{xcolor}
\definecolor{pastelgray}{rgb}{0.81, 0.81, 0.77}
\definecolor{beaublue}{rgb}{0.9, 0.9, 0.93}

\usepackage{orcidlink}
\usepackage{tikz,xcolor,hyperref}

\definecolor{lime}{HTML}{A6CE39}
\DeclareRobustCommand{\orcidicon}{
	\begin{tikzpicture}
	\draw[lime, fill=lime] (0,0) 
	circle [radius=0.16] 
	node[white] {{\fontfamily{qag}\selectfont \tiny ID}};
	\draw[white, fill=white] (-0.0625,0.095) 
	circle [radius=0.007];
	\end{tikzpicture}
	\hspace{-2mm}
}

\foreach \x in {A, ..., Z}{%
	\expandafter\xdef\csname orcid\x\endcsname{\noexpand\href{https://orcid.org/\csname orcidauthor\x\endcsname}{\noexpand\orcidicon}}
}

\makeatletter
\def\@bibdataout@aps{%
\immediate\write\@bibdataout{%
@CONTROL{%
apsrev41Control%
\longbibliography@sw{%
    ,author="08",editor="1",pages="1",title="0",year="1"%
    }{%
    ,author="08",editor="1",pages="1",title="",year="1"%
    }%
  }%
}%
\if@filesw \immediate \write \@auxout {\string \citation {apsrev41Control}}\fi
}
\makeatother

\newcolumntype{Y}{>{\centering\arraybackslash}X}

\begin{document}
\begin{CJK*}{UTF8}{gbsn}

\title{Structure evolution of ground and excited states in the exotic nucleus $^{22}$Al}

\author{Z. C. Xu (许志成)\,\orcidlink{0000-0001-5418-2717}}
\affiliation{Key Laboratory of Nuclear Physics and Ion-beam Application (MOE), Institute of Modern Physics, Fudan University, Shanghai 200433, China}
\affiliation{Shanghai Research Center for Theoretical Nuclear Physics, NSFC and Fudan University, Shanghai 200438, China}

\author{H. Y. Shang (商浩宇)\,\orcidlink{0009-0007-1253-4519}}
\affiliation{Key Laboratory of Nuclear Physics and Ion-beam Application (MOE), Institute of Modern Physics, Fudan University, Shanghai 200433, China}
\affiliation{Shanghai Research Center for Theoretical Nuclear Physics, NSFC and Fudan University, Shanghai 200438, China}

\author{S. M. Wang (王思敏)\,\orcidlink{0000-0002-8902-6842}}\email{Email: wangsimin@fudan.edu.cn}
\affiliation{Key Laboratory of Nuclear Physics and Ion-beam Application (MOE), Institute of Modern Physics, Fudan University, Shanghai 200433, China}
\affiliation{Shanghai Research Center for Theoretical Nuclear Physics, NSFC and Fudan University, Shanghai 200438, China}

\author{Y. G. Ma (马余刚)\,\orcidlink{0000-0002-0233-9900}}
\affiliation{School of Physics, East China Normal University, Shanghai 200241, China}
\affiliation{Key Laboratory of Nuclear Physics and Ion-beam Application (MOE), Institute of Modern Physics, Fudan University, Shanghai 200433, China}
\affiliation{Shanghai Research Center for Theoretical Nuclear Physics, NSFC and Fudan University, Shanghai 200438, China}


\begin{abstract}
Recent experimental studies on proton-rich nuclei in the $sd$ shell have revealed intriguing near-threshold phenomena, including exotic structures associated with mirror-symmetry breaking. In particular, a halo-like structure has been suggested for the $1^+$ state of $^{22}$Al based on the large isospin asymmetry observed in the $^{22}$Si/$^{22}$O mirror Gamow–Teller transitions. Recent mass measurements further indicate that the ground state of $^{22}$Al is weakly bound, with a single-proton separation energy of about 100 keV. To investigate how the continuum affects the structure and decay properties of this proton-dripline nucleus, we employ the state-of-the-art Gamow shell model. This approach utilizes valence-space effective interactions and operators derived from chiral forces. Our calculations identify the ground state of $^{22}$Al as a $4^+$ state, with a $3^+$ state as the first excitation. Despite their diffuse nature under weak binding, the Thomas-Ehrman shift for these states is found to be negligible due to their small $s$-wave components. In contrast, the excited $1_1^+$ state possesses a significantly larger $s$-wave component, resulting in a more pronounced halo-like structure.
\end{abstract}


\maketitle
\end{CJK*}

\section{Introduction}
As nuclei move away from the $\beta$-stability line, the growing imbalance between proton and neutron numbers enhances isospin-dependent effects. Near the driplines, mirror symmetry breaking (MSB) is significantly amplified \cite{Nazarewicz2025}, accompanied by the emergence of exotic structures \cite{Wylie2021,Yu2024,Hagen2012a} and rare decay modes \cite{Pfutzner2023,Blank2008_2,Webb2019,Wang2021}. In proton-rich nuclei, the Coulomb interaction pushes the proton Fermi surface and orbitals closer to the particle-emission threshold, making them particularly sensitive to continuum coupling~\cite{Nazarewicz1996,Michel2009,Michel2021,Hoffman2014,Hoffman2016,Miller2019,Xu2025}. Such continuum effects can substantially alter both single-particle orbitals \cite{Wang2019,Michel2023,Volya2005,Michel2009} and pairing correlations \cite{Grigorenko2002}. A well-known manifestation is the Thomas–Ehrman shift (TES), first observed in the $^{13}$N/$^{13}$C mirror pair \cite{Thomas1952,Ehrman1951} and later identified in dripline nuclei \cite{Wang2019,Stefan2014,Phillips2025,Zhang2023,Li2025dynamicTES,Zhang2025} and unbound systems \cite{Jin2021,Zhou2024,Xu2025a,Charity2023,Ma2025,Pfutzner2023,Zhou2022,Wang2025,Michel2021a,Kokkonen2025,XuXY2024,Fang2023}.



Among proton-rich isotopes, nuclei near $Z=14$ have recently attracted attention due to an unusual structural evolution driven by the proximity of the proton $s_{1/2}$ orbital to the continuum \cite{Jin2021,Lee2020,Xu2025a}. For instance, $^{22}$Si is weakly bound, with a two-proton separation energy of only about 200~keV, yet the $Z=14$ shell closure remains robust \cite{Xing2025,Phillips2025}. Significant isospin asymmetry has also been observed in excitation energies and $\beta$-decay Gamow–Teller strengths within the $^{22}$Si/$^{22}$O mirror pair, interpreted as evidence for a halo-like $1^+_1$ state in $^{22}$Al~\cite{Lee2020}. 

Nevertheless, the nature of the $^{22}$Al low-lying states remains controversial. While AME2020 suggested that $^{22}$Al is an unbound nucleus~\cite{AME2021}, recent high-precision mass measurements report weak binding with a single-proton separation energy around 100~keV~\cite{Campbell2024,Sun2024}. This weakly bound property raises the question of whether its ground state may exhibit a halo structure. Shell-model calculations predict low-lying $3^+$ and $4^+$ states with only small $s_{1/2}$ components~\cite{Campbell2024}, and the mirror energy difference (MED) of the separation energy along the $Z=13$ isotonic chain suggests that $^{22}$Al is unlikely to be a halo nucleus~\cite{Yu2024}. However, results from particle--rotor calculations propose that deformation of the $^{21}$Mg core may enhance the $s_{1/2}$ occupancy in the weakly bound $3^+$ state of $^{22}$Al, potentially modifying its near-threshold structure~\cite{Campbell2024}. Recent $\beta$-delayed $\alpha$-emission measurement suggests that a halo-like structure in the $^{22}$Al ground state is unlikely and indicates a $4^+$ assignment dominated by $d$-wave components~\cite{Jensen2026}.

To clarify these questions, we investigate the $^{22}$Al/$^{22}$F mirror pair using the state-of-the-art Gamow shell model (GSM)~\cite{Hu2020,Sun2017}, which derives its valence-space effective Hamiltonians and operators from chiral effective field theory (EFT) using many-body perturbation theory (MBPT)~\cite{Coraggio2020}, within the complex-energy Berggren basis \cite{Berggren1968}. Successfully applied to the spectroscopy and decay properties of dripline nuclei \cite{Xu2023,Zhang2022,Xu2025,Li2024}, this approach is suitable to describe $^{22}$Al, specifically regarding the complex interplay of isospin-dependent nuclear forces, many-body correlations, and continuum effects.


By analyzing Gamow–Teller and Fermi transitions, electric quadrupole ($E2$) transitions, and valence-nucleon densities, we reassess the configurations of $^{22}$Al and quantify mirror symmetry in both the ground and excited states. Our results provide a consistent description of the $^{22}$Al/$^{22}$F mirror pair, clarify the role of continuum coupling in their low-lying states, and shed light on the possible halo nature of $^{22}$Al.

\section{\label{sec:method} Method}

Starting from chiral two-nucleon and three-nucleon interactions, the intrinsic Hamiltonian of an $A$-body system can be written as  
\begin{equation}
    H = \sum_{i<j}^{A}\frac{(\bm{p}_{i}-\bm{p}_{j})^{2}}{2mA}
    + \sum_{i<j}^{A}v_{ij}^\text{NN}
    + \sum_{i<j<k}^{A}v_{ijk}^\text{3N},
    \label{eq1}
\end{equation}
where $\bm{p}$ denotes the nucleon momentum in the laboratory frame, $m$ is the nucleon mass, and $v^\text{NN}$ and $v^\text{3N}$ correspond to the two-nucleon force (2NF) and three-nucleon force (3NF), respectively. The chiral 2NF+3NF employed in this work is the interaction EM1.8/2.0~\cite{Hebeler2011,Machleidt2011}, which has been shown to provide a good description of many properties of nuclei up to medium masses~\cite{Stroberg2021,Miyagi2022}. In many-body calculations, the 3NF is normal-ordered with respect to a reference state, resulting in normal-ordered zero-, one-, and two-body terms, while the residual three-body term is neglected~\cite{Robert2012}.

To incorporate continuum effects at the basis level, we employ the Gamow–Hartree–Fock (GHF) method~\cite{Zhang2023,Hagen2006} to generate a Berggren basis, which treats bound, resonant, and non-resonant continuum states on an equal footing in the complex-momentum plane. Specifically, matrix elements of the chiral interaction and bare operators are first pre-calculated in a harmonic oscillator (HO) basis with $\hbar\omega = 16$ MeV, using 13 major shells (i.e., $e = 2n + l \leq e_{\rm max} = 12$) for two-nucleon forces and $e_{3\text{max}} = e_1+e_2+e_3 \leq 12$ for three-nucleon forces. These matrix elements are then transformed into the GHF basis~\cite{Zhang2023,Xu2023}, which subsequently defines the core and valence space.

For the nuclei of interest in this work, $^{16}$O is taken as the reference state (core). The GHF calculation for $^{16}$O yields bound $0d_{5/2}$ and resonant $0d_{3/2}$ orbits for both neutrons ($\nu$) and protons ($\pi$). The $\pi 1s_{1/2}$ orbit is resonant, whereas the $\nu 1s_{1/2}$ orbit remains bound. Consequently, the valence space for the many-body GSM is defined as follows: for neutron-rich nuclei, we use \{$\nu0d_{5/2}$, $\nu1s_{1/2}$, $\nu0d_{3/2}$ resonance plus continuum, $\pi0d_{5/2}$, $\pi1s_{1/2}$, $\pi0d_{3/2}$\}; for proton-rich nuclei, we use  \{$\nu0d_{5/2}$, $\nu1s_{1/2}$, $\nu0d_{3/2}$, $\pi0d_{5/2}$, $\pi1s_{1/2}$ resonance plus continuum, $\pi0d_{3/2}$ resonance plus continuum\}. 

The non-resonant continuum is included via  complex-momentum contours along the complex $k$-plane, chosen to ensure convergence. The contours are $k = 0 \rightarrow 0.35-0.20i \rightarrow 0.70 \rightarrow 4$ fm$^{-1}$ for the $\pi s_{1/2}$ partial wave, $k = 0 \rightarrow 0.60-0.20i \rightarrow 1.10 \rightarrow 4$ fm$^{-1}$ for the $\pi d_{3/2}$ wave, and $k = 0 \rightarrow 0.45-0.20i \rightarrow 0.70 \rightarrow 4$ fm$^{-1}$ for the $\nu d_{3/2}$ wave. Each contour is discretized with 35 scattering states. To manage computational cost, we restrict the number of valence particles in the scattering continuum to at most two~\cite{Sun2017,Hu2020,Michel2004}.

With the chosen valence space and reference state, the valence-particle effective Hamiltonian and effective operators are constructed consistently within MBPT~\cite{Xu2023}. Typically, the $\hat{S}$-box and $\hat{Q}$-box are evaluated up to third and second order, respectively, in the GHF basis~\cite{Sun2017,Hu2020}. In Ref.~\cite{Xu2023}, complex MBPT two-body matrix elements involving pole states (bound and resonant) were calculated up to third order. Here, we extend this by including third-order two-body matrix elements for all states, including those with non-resonant continuum configurations, thereby providing a more complete treatment of continuum coupling.


For comparison, we also perform standard shell-model (SM) calculations using the same chiral interaction EM1.8/2.0, but in a localized real-space Hartree-Fock basis that does not include explicit continuum effects. In both the GSM and SM calculations, we employ the effective Hamiltonian and effective operators derived from MBPT.

In this study, both Gamow–Teller (GT) and Fermi $\beta$ decays are investigated. The free-space bare GT and Fermi transition operators are given by~\cite{Brown1985}
\begin{equation}
    \mathcal{O}(\text{GT}_{\pm}) = \sum_{j}\sigma^j\tau^j_\pm,
    \qquad
    \mathcal{O}(\text{Fermi}_{\pm}) = \sum_{j}\tau^j_\pm,
\end{equation}
where $\sigma$ denotes the Pauli spin operator and $\tau_\pm=(\tau_x \pm i\tau_y)/2$ are the isospin raising and lowering operators corresponding to $\beta^\pm$ decay, respectively. The summation runs over all nucleons in the nucleus.

For GT transition calculations, a quenching factor is commonly introduced to account for missing correlations and to achieve agreement with experimental data~\cite{Brown1985,Towner1987}. For nuclei in the $sd$ and $pf$ shells, a typical phenomenological value is $q \approx 0.75$~\cite{Brown1985,Towner1987}. Within the chiral EFT framework, the renormalization of axial-vector transitions due to two-body currents (TBCs) can be treated explicitly~\cite{Gysbers2019,Hoferichter2020}. In this work, we adopt a density-dependent quenching factor derived in the Fermi-gas approximation~\cite{Hoferichter2020}. For a nuclear density of $\rho = 0.10~\mathrm{fm}^{-3}$, this corresponds to a quenching factor of $q = 0.78$, which is used in our GT calculations.

To investigate possible deformation effects, we also compute the $B(E2)$ values for the even-even nuclei within different model spaces. In line with our approach, the electromagnetic transition operator can be renormalized consistently~\cite{Xu2024,Xu2025,Fan2025}; consequently, no effective charges are employed in these calculations, following the prescription detailed in Refs.~\cite{Xu2025,Xu2024}. We also show the results of SM calculations using USDC interaction~\cite{Magilligan2020} with effective charges $e_p=1.5,\, e_n=0.5$.


\section{Results}

As shown in Table~\ref{tab:SeparationEnergy}, both our SM and GSM calculations reproduce the experimental single- and double-nucleon separation energies of the $^{22}$Si/$^{22}$O and $^{22}$Al/$^{22}$F mirror pairs. In particular, the ground states of $^{22}$Si and $^{22}$Al are predicted to lie very close to the two-proton and one-proton emission thresholds, respectively, highlighting their weakly-bound nature. This makes the detailed structure of the low-lying states in $^{22}$Al particularly interesting. In the mirror nucleus $^{22}$F, the $3^+$ state is located only 72~keV above the $4^+$ ground state, leading early studies to propose that $^{22}$Al might also exhibit a $3^+$ ground state with an enhanced $s_{1/2}$ occupation due to the TES~\cite{Blank1997b,Czajkowski1997}. Also, as introduced above, the core deformation in $^{21}$Mg could further increase this $s_{1/2}$ component~\cite{Campbell2024}, potentially giving rise to halo-like features. However, despite their proximity to the threshold, the calculated ground-state energies of $^{22}$Si and $^{22}$Al show only minor differences between SM and GSM, which suggests that continuum coupling plays a limited role in determining their structures in these states. 

\begin{figure}[h]
\includegraphics[width=1\columnwidth]{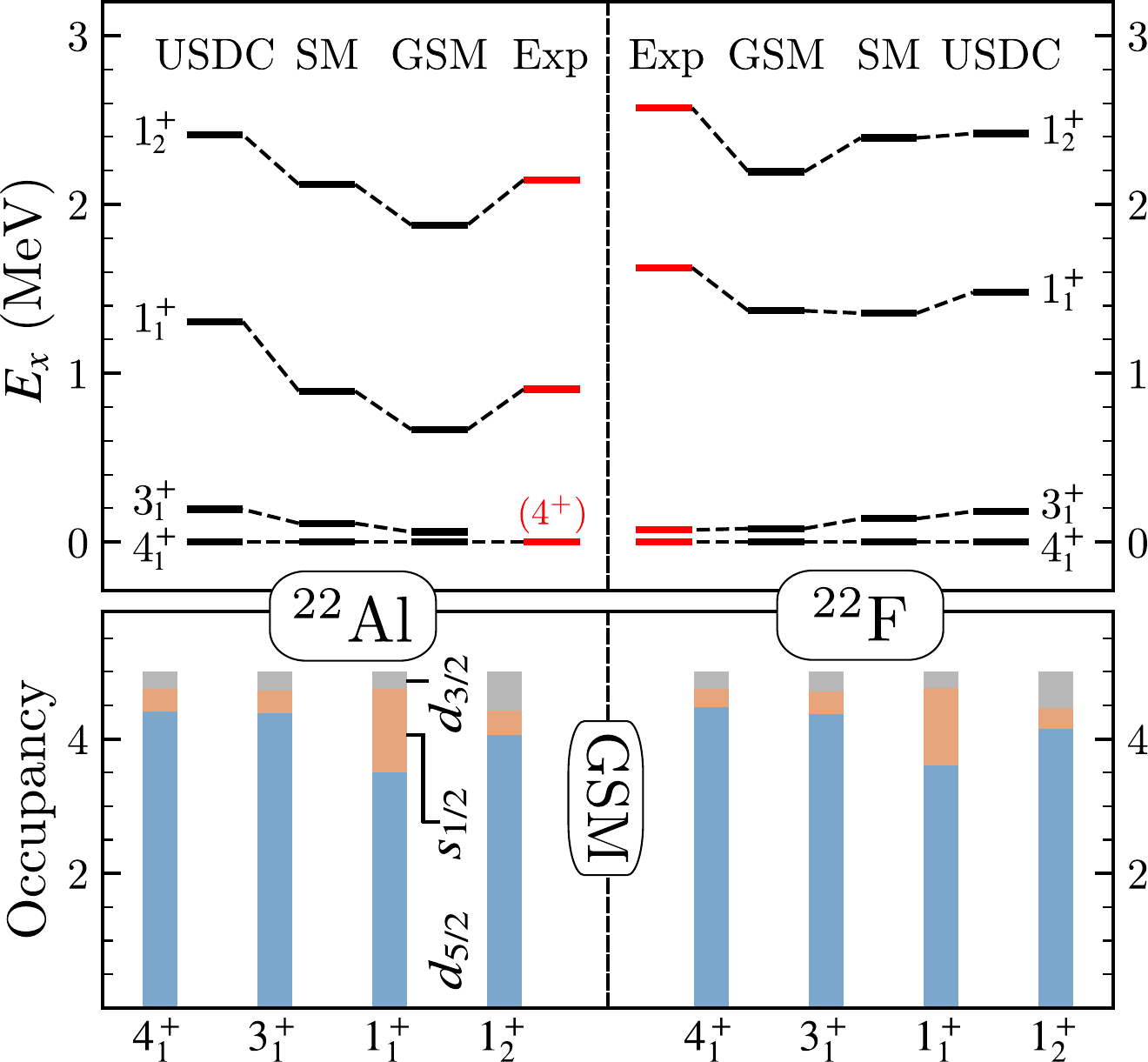}
\caption{The spectrum (upper) and occupation (lower) of $^{22}$Al/$^{22}$F mirror pairs. In the upper panel, the spectrum calculated by standard SM with USDC and EM1.8/2.0, and by GSM with EM1.8/2.0, compared with experimental data~\cite{Lee2020,Basunia2015}. In the lower panel, the valence proton/neutron occupation of states in $^{22}$Al/$^{22}$F mirror pairs calculated by GSM.}\label{spectrum}
\end{figure}

In detail, both our SM and GSM calculations consistently predict a $4^+$ ground state for $^{22}$Al, regardless of continuum effects (see Fig.~\ref{spectrum}). The first excited $3^+$ state is located just 111~keV (SM) or 60~keV (GSM) above the ground state. As shown in the lower panel of Fig.~\ref{spectrum}, the proton/neutron occupations for the $3^+$ and $4^+$ states in $^{22}$Al/$^{22}$F are very similar in the GSM results. The $s_{1/2}$ occupations (0.31 for $4^+$ and 0.41 for $3^+$) align closely with USD shell-model predictions~\cite{Campbell2024} and do not support the presence of a pronounced halo structure.

\begin{table}[]
\centering
\caption{The single- and double-nucleon separation energies of $^{22}$Si/$^{22}$O and $^{22}$Al/$^{22}$F mirror pairs from theoretical calculations and experimental data~\cite{Basunia2015,Campbell2024,Sun2024,Xing2025}. The SM and GSM results are calculated by standard SM and GSM with EM1.8/2.0, respectively.}~\label{tab:SeparationEnergy}
\begin{tabular*}{\hsize}{@{}@{\extracolsep{\fill}}ccccccc@{}c}
\toprule\toprule 
& & & USDC~\cite{Magilligan2020} &  SM  &  GSM  &  Exp &\\
\midrule
&  \multirow{4}{*}{$^{22}\text{Al}$}&$S_p$   & 0.092 & 0.339 & 0.200 & 0.1004(8)~\cite{Campbell2024} &\\
&                                   &        &  &     &  & 0.090(10)~\cite{Sun2024} &\\
&                                   &$S_{2p}$& 3.187 & 2.936 & 2.909 & 3.336(1)~\cite{Campbell2024} &\\
&                                   &        &   &    &  & 3.326(10)~\cite{Sun2024} &\\
\midrule
& \multirow{2}{*}{$^{22}\text{Si}$} &$S_p$   & 1.527 & 0.985 & 1.044 & 1.412(114) \cite{Xing2025} &\\
&                                   &$S_{2p}$& 0.280 & 0.089 & 0.004 & 0.229(54) \cite{Xing2025} &\\
\midrule
&  \multirow{2}{*}{$^{22}\text{F}$} &$S_n$   & 5.09 & 5.432 & 5.334 &  5.230(13)~\cite{Basunia2015} &\\
&                                   &$S_{2n}$& 12.898 &12.952 &12.947 & 13.332(12)~\cite{Basunia2015}  &\\
\midrule
& \multirow{2}{*}{$^{22}\text{O}$}  &$S_n$   & 6.903 & 6.578 & 6.568 &  6.850(58)~\cite{Basunia2015}  &\\
&                                   &$S_{2n}$& 10.691&10.736 &10.675 & 10.656(57)~\cite{Basunia2015}  &\\
\bottomrule
\end{tabular*}
\end{table}

Mirror symmetry in the low-lying structure is further examined by studying the $\beta$ decay of $^{22}$Al/$^{22}$F to their daughter nuclei $^{22}$Mg/$^{22}$Ne. Table~\ref{tab:beta} shows that both the calculated excitation energies and $\log ft$ values agree well with experimental data~\cite{Wu2021,Davids1974}. The observed mirror symmetry in final states energies and $\log ft$ values, combined with the small branching ratio to the $3^+$ level, provides strong support for a $4^+$ ground state in $^{22}$Al. Since $\beta$ decay is sensitive to the microscopic structure of the mother and daughter nuclei, the good agreement of mirror $\log ft$ values indicates similar ground-state configurations in $^{22}$Al and $^{22}$F. Our SM and GSM calculations consistently reproduce the spectra and transition strengths, validating the calculated configurations.





\begin{table}[t]
\centering
\caption{The $\beta$-decay transition $\log ft$ values and branching ratios $I\beta^{\pm}$ for the mirror nuclei $^{22}\mathrm{Al}$ and $^{22}\mathrm{F}$ decaying into the $3^+$, $4^+$, and $5^+$ states of their mirror daughters $^{22}\mathrm{Mg}$ and $^{22}\mathrm{Ne}$. The SM and GSM results are calculated by standard SM and GSM with EM1.8/2.0, respectively. Experimental data for $^{22}\mathrm{Mg}$ are taken from Ref.~\cite{Wu2021} and for $^{22}\mathrm{Ne}$ from Ref.~\cite{Davids1974}.}
\label{tab:beta}

\setlength{\tabcolsep}{4pt} 
\begin{tabular}{ccclll}
\toprule\toprule
Decay & $J_f^\pi$ &  & $E_f$ (MeV) & $\log ft$ & $I\beta^{\pm}$ (\%) \\
\midrule

\multirow{9}{*}{$^{22}\mathrm{Al}\to{}^{22}\mathrm{Mg}$}
& \multirow{3}{*}{$4^+$}
& Exp & 5.294(1) & 4.93(7)  & 26.91(390) \\
& & SM  & 5.569    & 4.89     & -- \\
& & GSM & 5.527    & 4.86     & -- \\
\cmidrule(lr){2-6}

& \multirow{3}{*}{$3^+$}
& Exp & 5.453(1) & 5.43(17) & 8.41(325) \\
& & SM  & 5.582    & 5.14     & -- \\
& & GSM & 5.537    & 5.09     & -- \\
\cmidrule(lr){2-6}

& \multirow{3}{*}{$5^+$}
& Exp & 7.135(7) & 4.70(1)  & 19.75(26) \\
& & SM  & 7.145    & 4.74     & -- \\
& & GSM & 7.095    & 4.71     & -- \\
\midrule

\multirow{9}{*}{$^{22}\mathrm{F}\to{}^{22}\mathrm{Ne}$}
& \multirow{3}{*}{$4^+$}
& Exp & 5.5235(7) & 4.802(16) & 53.9(15) \\
& & SM  & 5.776     & 5.37      & -- \\
& & GSM & 5.740     & 5.20      & -- \\
\cmidrule(lr){2-6}

& \multirow{3}{*}{$3^+$}
& Exp & 5.6413(8) & 5.274(22) & 16.4(7) \\
& & SM  & 5.871     & 5.30      & -- \\
& & GSM & 5.874     & 5.35      & -- \\
\cmidrule(lr){2-6}

& \multirow{3}{*}{$5^+$}
& Exp & 7.4236(9) & 4.725(25) & 8.7(4) \\
& & SM  & 7.478     & 4.58      & -- \\
& & GSM & 7.471     & 4.55      & -- \\
\bottomrule
\end{tabular}
\end{table}

\begin{figure}[h]
\includegraphics[width=1\columnwidth]{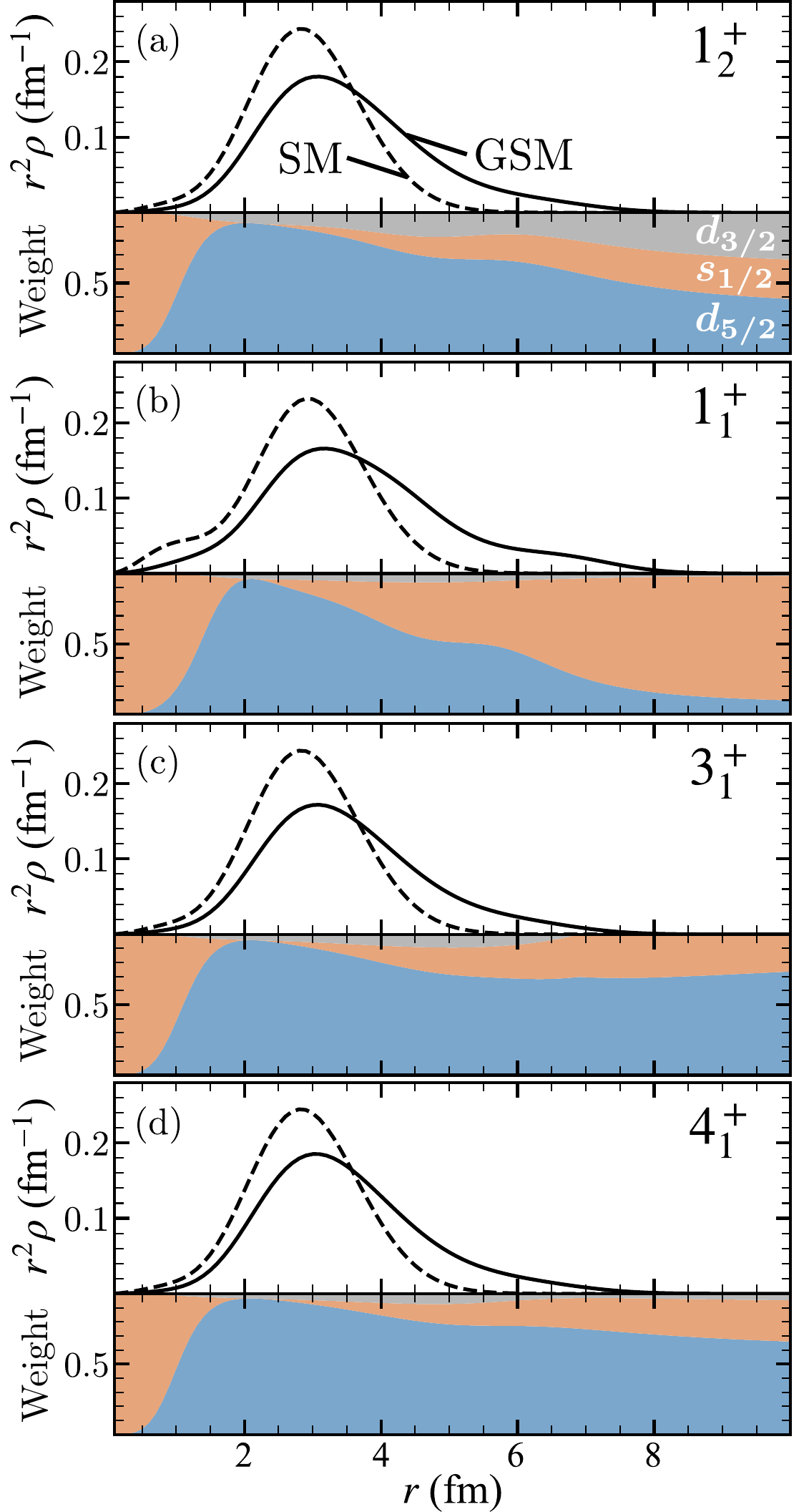}
\caption{Valence proton density $r^2\rho$ of $^{22}$Al with respect to the $^{16}$O core, obtained from SM (dashed lines) and GSM (solid lines) calculations using the chiral interaction EM1.8/2.0. The contributions of the $d_{3/2}$, $s_{1/2}$, and $d_{5/2}$ partial-wave proton densities in the GSM calculations are shown in the lower part of each panel and are indicated by the gray, orange, and blue bands, respectively. The results for the $1^+_2$, $1^+_1$, $3^+_1$, and $4^+_1$ states are displayed in panels (a), (b), (c), and (d), respectively.}\label{density}
\end{figure}

To visualize structural evolution among the low-lying states of $^{22}$Al and to demonstrate the role of continuum coupling, we computed the valence nucleon radial densities (relative to the $^{16}$O core) for the $4^+_1$, $3^+_1$, $1^+_1$, and $1^+_2$ states within both SM and GSM. Although the $1^+_1$ and $1^+_2$ states are particle-unbound, which indicates that their wave functions have a non-trivial asymptotic form, and a standard root-mean-square radius cannot be used to define halo structure unambiguously. Nevertheless, a diffuse, halo-like character can be inferred from the density, continuum coupling, and an enhancement in spatial extent, as discussed in Ref.~\cite{Lin2025}. As shown in Fig.~\ref{density}, the GSM densities (solid lines) exhibit reduced interior amplitudes and more diffuse surface distributions compared to their SM counterparts (dashed lines), a direct signature of the weakly bound or unbound nature of these states. Thus, only the GSM calculation that includes continuum coupling yields the correct asymptotic tail of the wave function. As discussed in Refs.~\cite{Michel2009,Xu2025}, such a tail is essential for describing observables of dripline nuclei.

Among the four states, the $1^+_1$ state displays a distinctly extended radial tail, characteristic of a halo-like configuration. In contrast, the other three states present relatively compact density profiles. The halo-like structure in the $1^+_1$ state emerges from strong coupling to the $s_{1/2}$ continuum, as illustrated in the partial-wave decomposition of the proton densities (lower sections of each panel in Fig.~\ref{density}). Across all states, the $d_{5/2}$ component dominates the interior region. However, in the $1^+_1$ state, the long-range tail arises almost entirely from the $s_{1/2}$ partial wave, which shows significant amplitude at large radii, supported by non-negligible scattering $s$-wave contributions.

For the $1^+_2$ state, a noticeable $d_{3/2}$ component produces a moderate enhancement near the nuclear surface. In the $3^+_1$ and $4^+_1$ states, the asymptotic density remains governed by the $d_{5/2}$ wave. Although the $3^+_1$ state possesses slightly larger $s_{1/2}$ and $d_{3/2}$ components near the surface compared to the $4^+_1$ state.



As deformation has been proposed as a possible mechanism to increase the $s_{1/2}$ occupation in the $3^+$ state of $^{22}$Al~\cite{Campbell2024}---though this effect cannot be fully captured in current theoretical frameworks---we examine the $B(E2)$ values of other $A=22$ nuclei, since $B(E2)$ values serve as a good indicator of quadrupole collectivity. Calculations are performed in both the $sd$ and $sdpf$ model spaces to further assess the influence of cross-shell excitations and deformation. As shown in Table~\ref{Tab:be2}, the calculated $E(2^+_1)$ energies and $B(E2;0^+_1\rightarrow 2^+_1)$ values for the $^{22}$Mg/$^{22}$Na/$^{22}$Ne isobaric triplet and the $^{22}$Si/$^{22}$O mirror pair are in good agreement with experimental data and remain stable across the two valence spaces. However, the $B(E2)$ values obtained with EM1.8/2.0, where no effective charges are employed, are smaller than those obtained with USDC. As discussed in Refs.~\cite{Stroberg2022,Sun2025}, excitations outside the valence space are essential to account for the missing $B(E2)$ strength. Moreover, the $B(E2)$ values for this isobaric triplet deviate from a linear trend~\cite{Backes2025}, indicating a complex pattern of isospin-symmetry breaking.

\begin{table}[]
\centering
\caption{The $E(2^+_1)$ (MeV) and $B(E2;0^+_1\rightarrow2^+_1)$ ($e^2\,\mathrm{fm}^4$) values for the $^{22}$Mg/$^{22}$Na/$^{22}$Ne isobaric triplet and the $^{22}$Si/$^{22}$O mirror pair. For $^{22}$Na, $E(2^+_{1})$ corresponds to the energy gap between the isobaric analog states, $E(2^+_{\mathrm{IAS}}-0^+_{\mathrm{IAS}})$. The SM and GSM results are obtained with the chiral interaction EM1.8/2.0 using different valence spaces. SM results obtained with the USDC interaction are also shown for comparison. Experimental data are taken from Refs.~\cite{Phillips2025,Stanoiu2004,Basunia2015}.\label{Tab:be2}}
\begin{tabular*}{\hsize}{@{}@{\extracolsep{\fill}}lccccccc@{}}
\toprule\toprule
& & & \multicolumn{2}{c}{$sd$-space} & \multicolumn{2}{c}{$sdpf$-space}\\
\cmidrule(lr){4-5} \cmidrule(lr){6-7} 

& & USDC & SM & GSM & SM & GSM & Exp\\ 
\midrule
\multirow{2}{*}{$^{22}$Mg}
 & $E(2^+_1)$& 1.259 & 1.033 & 1.136 & 1.031 & 1.134 & 1.24702(3)~\cite{Basunia2015}     \\
 & $B(E2)$   & 399   & 237   & 221   & 237   & 220   & 476(201)~\cite{Basunia2015}     \\
\midrule
\multirow{2}{*}{$^{22}$Na}
 & $E(2^+_1)$& 1.291 &1.116& 1.159 & 1.116 & 1.161 & 1.2948~\cite{Basunia2015} \\
 & $B(E2)$   & 348   & 182 & 190  & 182  & 193  & 275(92)~\cite{Basunia2015}  \\
\midrule
\multirow{2}{*}{$^{22}$Ne}
 & $E(2^+_1)$& 1.262 & 1.101 & 1.191 & 1.098 & 1.119 & 1.274537(7)~\cite{Basunia2015} \\
 & $B(E2)$   & 301   & 135   & 134   & 135   & 135   & 234(3)~\cite{Basunia2015}  \\
\midrule
\multirow{2}{*}{$^{22}$Si}
 & $E(2^+_1)$& 2.769 & 1.955 & 1.644 & 1.930 & 1.637 & 2.352(55)~\cite{Phillips2025}\\
 & $B(E2)$   & 223   & 167   & 206   & 170   & 206   & --\\
\midrule
\multirow{2}{*}{$^{22}$O}
 & $E(2^+_1)$& 3.005 & 2.651 & 2.670 & 2.630 & 2.530 & 3.199(8)~\cite{Stanoiu2004}\\
 & $B(E2)$   & 24    & 2     & 2     & 2     & 2     & 22(9)~\cite{Basunia2015}\\
\bottomrule
\bottomrule
\end{tabular*}
\end{table}

Interestingly, while the ground states of the $^{22}$Si/$^{22}$O mirror pair show good symmetry in our predictions --- including energy and configurations~\cite{Xing2025,Xu2023} --- their $2^+_1$ excited states exhibit markedly different excitation energies (see Table~\ref{Tab:be2}). This contrasts with predictions from the USDC interaction, which incorporates isospin-breaking effects phenomenologically through fitted matrix elements~\cite{Magilligan2020}. In Ref.~\cite{Phillips2025}, the spectra were reproduced by an adjusted interaction (USDP) that lowers the $1s_{1/2}$ single-particle energy by 700 keV relative to USDC. Our calculations show that a large MED for the $2^+_1$ states is already pronounced in the SM with the EM1.8/2.0 interaction, since the renormalization partly accounts for the contributions from high-momentum space. This MED is further enhanced by continuum coupling. This behavior agrees with recent experimental measurements, which report an MED of 0.847(56) MeV for this pair~\cite{Phillips2025}.

Furthermore, similar to the case of $^{36}$Ca demonstrated in our previous study~\cite{Xu2025}, the weakly-bound ground state and unbound $2^+_1$ state in $^{22}$Si lead to an enhanced $B(E2)$ value when comparing the SM and GSM results (Table~\ref{Tab:be2}). These findings underscore the crucial role of the TES and continuum coupling in shaping the excited-state structure of $^{22}$Si by the diffused asymptotic wave functions. Consequently, $^{22}$Si provides a compelling test case for experimentally probing the interplay between continuum effects and the $Z=14$ shell structure.



\section{Summary}

In this work, we investigate the low-lying structure of the weakly bound nucleus $^{22}$Al and its mirror partner $^{22}$F using the Gamow Shell Model (GSM) with effective interactions derived from many-body perturbation theory. This approach self-consistently incorporates continuum coupling and realistic interactions. Our calculations successfully reproduce the separation energies and decay properties of the $^{22}$Al/$^{22}$F mirror pair and other nuclei in the $A=22$ isobaric chain.

Both the GSM and calculations without the continuum predict a $4^+$ ground state for $^{22}$Al, with a nearby $3^+_1$ excited state. These states possess only a small $s_{1/2}$ component, indicating that continuum coupling plays a minor role and that neither state exhibits a pronounced halo structure, despite the system's weak binding. The $4^+$ ground-state assignment is further supported by the $\beta$-decay properties, which---due to angular momentum coupling selection rules---also rule out a halo configuration for the ground state. In contrast, the excited $1^+_1$ state displays clear signatures of a halo-like structure. It is characterized by a significantly enhanced $s_{1/2}$ component, a markedly extended density tail, and strong coupling to the continuum.

Additionally, our framework provides predictions for the ground- and excited-state properties of $^{22}$Si, such as energies and $E2$ transition strengths, offering a basis for further investigation into continuum coupling and shell evolution here.

{\it Acknowledgments.} --- This material is based upon work supported by the National Natural Science Foundation of China under Contract No.\,12347106, No.\,12547102, and No.\,12447122; the National Key Research and Development Program (MOST 2023YFA1606404 and MOST 2022YFA1602303);  the China Postdoctoral Science Foundation under Contract No.\,2024M760489. The authors thank W. Nazarewicz for fruitful discussions.

{\it Data availability.} --- The data are not publicly available. The data are available from the authors upon reasonable request.

\bibliography{references}

@article{Backes2025,
  title = {Quadrupole strength in isobaric triplets},
  author = {Backes, B. C. and Dobaczewski, J. and Muir, D. and Nazarewicz, W. and Reinhard, P.-G. and Bentley, M. A. and Wadsworth, R.},
  journal = {Phys. Rev. C},
  volume = {112},
  issue = {6},
  pages = {064311},
  numpages = {10},
  year = {2025},
  month = {Dec},
  publisher = {American Physical Society},
  doi = {10.1103/5wkc-7gmf},
  url = {https://link.aps.org/doi/10.1103/5wkc-7gmf}
}

@article{Basunia2015,
  title = {Nuclear {{Data Sheets}} for {{A}} = 22},
  author = {Basunia, M. Shamsuzzoha},
  year = 2015,
  journal = {Nuclear Data Sheets},
  volume = {127},
  pages = {69--190},
  issn = {0090-3752},
  doi = {10.1016/j.nds.2015.07.002},
}

@article{Berggren1968,
author = {Berggren, T.},
doi = {10.1016/0375-9474(68)90593-9},
isbn = {2003013555},
issn = {03759474},
journal = {Nucl. Phys. A},
number = {2},
pages = {265--287},
title = {On the use of resonant states in eigenfunction expansions of scattering and reaction amplitudes},
volume = {109},
year = {1968}
}

@article{Blank1997b,
  title = {The Spectroscopy of {{$^{22}$Al}}: A $\beta p$, $\beta 2p$ and $\beta \alpha$ Emitter},
  author = {Blank, B. and Boué, F. and Andriamonje, S. and Czajkowski, S. and Del Moral, R. and Dufour, J.P. and Fleury, A. and Pourre, P. and Pravikoff, M.S. and Orr, N.A. and Schmidt, K.-H. and Hanelt, E.},
  date = {1997},
  journal = {Nuclear Physics A},
  volume = {615},
  number = {1},
  pages = {52--68},
  issn = {0375-9474},
  doi = {10.1016/S0375-9474(96)00483-6},
  url = {https://www.sciencedirect.com/science/article/pii/S0375947496004836},
}

@article{Blank2008_2,
title = "Nuclear structure at the proton drip line: Advances with nuclear decay studies",
journal = "Prog. Part. Nucl. Phys.",
volume = "60",
number = "2",
pages = "403--483",
year = "2008",
issn = "0146-6410",
doi = "10.1016/j.ppnp.2007.12.001",
author = "B. Blank and M. J. G. Borge"
}

@article{Brown1985,
title = {Experimental and theoretical Gamow-Teller beta-decay observables for the sd-shell nuclei},
journal = {At. Data Nucl. Data Tables},
volume = {33},
number = {3},
pages = {347-404},
year = {1985},
issn = {0092-640X},
doi = {https://doi.org/10.1016/0092-640X(85)90009-9},
url = {https://www.sciencedirect.com/science/article/pii/0092640X85900099},
author = {B.A. Brown and B.H. Wildenthal},
}

@article{Campbell2024,
  title = {Precision Mass Measurement of the Proton Dripline Halo Candidate $^{22}\mathrm{Al}$},
  author = {Campbell, S. E. and Bollen, G. and Brown, B. A. and Dockery, A. and Ireland, C. M. and Minamisono, K. and Puentes, D. and Rickey, B. J. and Ringle, R. and Yandow, I. T. and Fossez, K. and Ortiz-Cortes, A. and Schwarz, S. and Sumithrarachchi, C. S. and Villari, A. C. C.},
  journal = {Phys. Rev. Lett.},
  volume = {132},
  issue = {15},
  pages = {152501},
  numpages = {7},
  year = {2024},
  month = {Apr},
  publisher = {American Physical Society},
  doi = {10.1103/PhysRevLett.132.152501},
  url = {https://link.aps.org/doi/10.1103/PhysRevLett.132.152501}
}

@ARTICLE{Coraggio2020,
TITLE={Perturbative Approach to Effective Shell-Model Hamiltonians and Operators},      
AUTHOR={Coraggio, Luigi and Itaco, Nunzio},   
JOURNAL={Front. Phys.},      
VOLUME={8},      
pages = {345},
YEAR={2020},      
URL={https://www.frontiersin.org/article/10.3389/fphy.2020.00345},       
DOI={10.3389/fphy.2020.00345},      
ISSN={2296-424X}
}

@article{Charity2023,
  title = {Strong Evidence for $^{9}\mathrm{N}$ and the Limits of Existence of Atomic Nuclei},
  author = {Charity, R. J. and Wylie, J. and Wang, S. M. and Webb, T. B. and Brown, K. W. and Cerizza, G. and Chajecki, Z. and Elson, J. M. and Estee, J. and Hoff, D. E. M. and Kuvin, S. A. and Lynch, W. G. and Manfredi, J. and Michel, N. and McNeel, D. G. and Morfouace, P. and Nazarewicz, W. and Pruitt, C. D. and Santamaria, C. and Sweany, S. and Smith, J. and Sobotka, L. G. and Tsang, M. B. and Wuosmaa, A. H.},
  journal = {Phys. Rev. Lett.},
  volume = {131},
  issue = {17},
  pages = {172501},
  numpages = {7},
  year = {2023},
  month = {Oct},
  publisher = {American Physical Society},
  doi = {10.1103/PhysRevLett.131.172501},
  url = {https://link.aps.org/doi/10.1103/PhysRevLett.131.172501}
}

@article{Czajkowski1997,
  title = {Beta-p,-2p,-Alpha Spectroscopy of 22,23,24si and 22al},
  author = {Czajkowski, S. and Andriamonje, S. and Blank, B. and Bou{\'e}, F. and Del Moral, R. and Dufour, J.P. and Fleury, A. and Hanelt, E. and Orr, N.A. and Pourre, P. and Pravikoff, M.S. and Schmidt, K.-H.},
  year = {1997},
  journal = {Nuclear Physics A},
  volume = {616},
  number = {1},
  pages = {278--285},
  issn = {0375-9474},
  doi = {10.1016/S0375-9474(97)00098-5}
}

@article{Davids1974,
  title = {$\ensuremath{\beta}$ decay of $^{22}\mathrm{F}$},
  author = {Davids, Cary N. and Goosman, David R. and Alburger, David E. and Gallmann, A. and Guillaume, G. and Wilkinson, D. H. and Lanford, W. A.},
  journal = {Phys. Rev. C},
  volume = {9},
  issue = {1},
  pages = {216--226},
  numpages = {0},
  year = {1974},
  month = {Jan},
  publisher = {American Physical Society},
  doi = {10.1103/PhysRevC.9.216},
  url = {https://link.aps.org/doi/10.1103/PhysRevC.9.216}
}

@article{Ehrman1951,
  title = {On the Displacement of Corresponding Energy Levels of $^{}\mathrm{C}^{13}$ and $^{}\mathrm{N}^{13}$},
  author = {Ehrman, J. B.},
  journal = {Phys. Rev.},
  volume = {81},
  issue = {3},
  pages = {412--416},
  numpages = {0},
  year = {1951},
  month = {Feb},
  publisher = {American Physical Society},
  doi = {10.1103/PhysRev.81.412}
}

@article{Fan2025,
  title = {Ab Initio Calculations of the Highest-Multipole Electromagnetic Transition Ever Observed in Nuclei},
  author = {Fan, Si Qin and Yuan, Qi and Xu, Fu-Rong and Walker, Philip Malzard},
  year = 2025,
  month = nov,
  journal = {Nucl. Sci. Tech.},
  volume = {36},
  number = {11},
  pages = {222},
  issn = {1001-8042, 2210-3147},
  doi = {10.1007/s41365-025-01812-2},
  urldate = {2026-01-06},
  langid = {english},
}

@article{Fang2023,
author = {Fang, D. Q. and Hua, H. and Ma, Y. G. and Wang, S. M.},
title = {Exploring the Edge of Nuclear Stability on the Proton-Rich Side},
journal = {Nucl. Phys. News},
volume = {33},
number = {2},
pages = {11--16},
year = {2023},
publisher = {Taylor \& Francis},
doi = {10.1080/10619127.2023.2168911},
URL = {https://doi.org/10.1080/10619127.2023.2168911}
}

@article{Grigorenko2002,
  title = {Two-Proton Widths of $^{12}\mathrm{O}$, $^{16}\mathrm{Ne}$, and Three-Body Mechanism of Thomas-Ehrman Shift},
  author = {Grigorenko, L. V. and Mukha, I. G. and Thompson, I. J. and Zhukov, M. V.},
  journal = {Phys. Rev. Lett.},
  volume = {88},
  issue = {4},
  pages = {042502},
  numpages = {4},
  year = {2002},
  month = {Jan},
  publisher = {American Physical Society},
  doi = {10.1103/PhysRevLett.88.042502}
}

@article{Gysbers2019,
  title = {Discrepancy between experimental and theoretical $\beta$-decay rates resolved from first principles},
  author = {Gysbers, P. and Hagen, G. and Holt, J. D. and Jansen, G. R. and Morris, T. D. and Navrátil, P. and Papenbrock, T. and Quaglioni, S. and Schwenk, A. and Stroberg, S. R. and Wendt, K. A.},
  journal = {Nat. Phys.},
  volume = {15},
  issue = {5},
  pages = {428},
  numpages = {4},
  year = {2019},
  month = {May},
  doi = {10.1038/s41567-019-0450-7},
  url = {https://doi.org/10.1038/s41567-019-0450-7}
}

@article{Hagen2006,
  title = {Gamow shell model and realistic nucleon-nucleon interactions},
  author = {Hagen, G. and Hjorth-Jensen, M. and Michel, N.},
  journal = {Phys. Rev. C},
  volume = {73},
  issue = {6},
  pages = {064307},
  numpages = {13},
  year = {2006},
  month = {Jun},
  publisher = {American Physical Society},
  doi = {10.1103/PhysRevC.73.064307},
  url = {https://link.aps.org/doi/10.1103/PhysRevC.73.064307}
}

@article{Hagen2012a,
  title = {Continuum Effects and Three-Nucleon Forces in Neutron-Rich Oxygen Isotopes},
  author = {Hagen, G. and Hjorth-Jensen, M. and Jansen, G. R. and Machleidt, R. and Papenbrock, T.},
  journal = {Phys. Rev. Lett.},
  volume = {108},
  issue = {24},
  pages = {242501},
  numpages = {4},
  year = {2012},
  month = {Jun},
  publisher = {American Physical Society},
  doi = {10.1103/PhysRevLett.108.242501},
  url = {https://link.aps.org/doi/10.1103/PhysRevLett.108.242501}
}

@article{Hebeler2011,
  title = {Improved nuclear matter calculations from chiral low-momentum interactions},
  author = {Hebeler, K. and Bogner, S. K. and Furnstahl, R. J. and Nogga, A. and Schwenk, A.},
  journal = {Phys. Rev. C},
  volume = {83},
  issue = {3},
  pages = {031301(R)},
  numpages = {5},
  year = {2011},
  month = {Mar},
  publisher = {American Physical Society},
  doi = {10.1103/PhysRevC.83.031301},
  url = {https://link.aps.org/doi/10.1103/PhysRevC.83.031301}
}

@article{Hoferichter2020,
  title = {Coherent elastic neutrino-nucleus scattering: EFT analysis and nuclear responses},
  author = {Hoferichter, Martin and Men\'endez, Javier and Schwenk, Achim},
  journal = {Phys. Rev. D},
  volume = {102},
  issue = {7},
  pages = {074018},
  numpages = {30},
  year = {2020},
  month = {Oct},
  publisher = {American Physical Society},
  doi = {10.1103/PhysRevD.102.074018},
  url = {https://link.aps.org/doi/10.1103/PhysRevD.102.074018}
}

@article{Hoffman2014,
  title = {Neutron $s$ states in loosely bound nuclei},
  author = {Hoffman, C. R. and Kay, B. P. and Schiffer, J. P.},
  journal = {Phys. Rev. C},
  volume = {89},
  issue = {6},
  pages = {061305},
  numpages = {5},
  year = {2014},
  month = {Jun},
  publisher = {American Physical Society},
  doi = {10.1103/PhysRevC.89.061305},
  url = {https://link.aps.org/doi/10.1103/PhysRevC.89.061305}
}

@article{Hoffman2016,
  title = {Ordering of the $0{d}_{5/2}$ and $1{s}_{1/2}$ proton levels in light nuclei},
  author = {Hoffman, C. R. and Kay, B. P. and Schiffer, J. P.},
  journal = {Phys. Rev. C},
  volume = {94},
  issue = {2},
  pages = {024330},
  numpages = {8},
  year = {2016},
  month = {Aug},
  publisher = {American Physical Society},
  doi = {10.1103/PhysRevC.94.024330},
  url = {https://link.aps.org/doi/10.1103/PhysRevC.94.024330}
}

@article{Hu2020,
title = {An ab-initio Gamow shell model approach with a core},
journal = {Phys. Lett. B},
volume = {802},
pages = {135206},
year = {2020},
issn = {0370-2693},
doi = {https://doi.org/10.1016/j.physletb.2020.135206},
url = {https://www.sciencedirect.com/science/article/pii/S0370269320300101},
author = {B. S. Hu and Q. Wu and J. G. Li and Y. Z. Ma and Z. H. Sun and N. Michel and F. R. Xu}
}

@misc{Jensen2026,
  title = {A Ground State {\textsuperscript{22}}{{Al}} Halo Is Unlikely},
  author = {Jensen, E. A. M. and Nielsen, J. S. and Johansson, B. S. O. and Adams, A. and Dopfer, J. and Sumithrarachchi, C. S. and Sun, L. J. and Weghorn, L. E. and Wheeler, T. and Wrede, C. and Borge, M. J. G. and Tengblad, O. and Madurga, M. and Jonson, B. and Riisager, K. and Fynbo, H. O. U.},
  year = 2026,
  month = jan,
  eprint = {2601.03961},
  archiveprefix = {arXiv},
}

@article{Jin2021,
  title = {First Observation of the Four-Proton Unbound Nucleus $^{18}\mathrm{Mg}$},
  author = {Jin, Y. and Niu, C. Y. and Brown, K. W. and Li, Z. H. and Hua, H. and Anthony, A. K. and Barney, J. and Charity, R. J. and Crosby, J. and Dell'Aquila, D. and Elson, J. M. and Estee, J. and Ghazali, M. and Jhang, G. and Li, J. G. and Lynch, W. G. and Michel, N. and Sobotka, L. G. and Sweany, S. and Teh, F. C. E. and Thomas, A. and Tsang, C. Y. and Tsang, M. B. and Wang, S. M. and Wu, H. Y. and Yuan, C. X. and Zhu, K.},
  journal = {Phys. Rev. Lett.},
  volume = {127},
  issue = {26},
  pages = {262502},
  numpages = {6},
  year = {2021},
  month = {Dec},
  publisher = {American Physical Society},
  doi = {10.1103/PhysRevLett.127.262502},
  url = {https://link.aps.org/doi/10.1103/PhysRevLett.127.262502}
}

@article{Kokkonen2025,
  title = {New Proton Emitter {{188At}} Implies an Interaction Unprecedented in Heavy Nuclei},
  author = {Kokkonen, Henna and Auranen, Kalle and Siwach, Pooja and others},
  year = 2025,
  month = may,
  journal = {Nature Communications},
  volume = {16},
  number = {1},
  pages = {4985},
  issn = {2041-1723},
  doi = {10.1038/s41467-025-60259-6},
  urldate = {2026-01-05},
  langid = {english}
}

@article{Lee2020,
  title = {Large Isospin Asymmetry in $^{22}\mathrm{Si}/^{22}\mathrm{O}$ Mirror Gamow-Teller Transitions Reveals the Halo Structure of $^{22}\mathrm{Al}$},
  author = {Lee, J. and Xu, X. X. and Kaneko, K. and Sun, Y. and Lin, C. J. and Sun, L. J. and Liang, P. F. and Li, Z. H. and Li, J. and Wu, H. Y. and Fang, D. Q. and Wang, J. S. and Yang, Y. Y. and Yuan, C. X. and Lam, Y. H. and Wang, Y. T. and Wang, K. and Wang, J. G. and Ma, J. B. and Liu, J. J. and Li, P. J. and Zhao, Q. Q. and Yang, L. and Ma, N. R. and Wang, D. X. and Zhong, F. P. and Zhong, S. H. and Yang, F. and Jia, H. M. and Wen, P. W. and Pan, M. and Zang, H. L. and Wang, X. and Wu, C. G. and Luo, D. W. and Wang, H. W. and Li, C. and Shi, C. Z. and Nie, M. W. and Li, X. F. and Li, H. and Ma, P. and Hu, Q. and Shi, G. Z. and Jin, S. L. and Huang, M. R. and Bai, Z. and Zhou, Y. J. and Ma, W. H. and Duan, F. F. and Jin, S. Y. and Gao, Q. R. and Zhou, X. H. and Hu, Z. G. and Wang, M. and Liu, M. L. and Chen, R. F. and Ma, X. W.},
  collaboration = {RIBLL Collaboration},
  journal = {Phys. Rev. Lett.},
  volume = {125},
  issue = {19},
  pages = {192503},
  numpages = {6},
  year = {2020},
  month = {Nov},
  publisher = {American Physical Society},
  doi = {10.1103/PhysRevLett.125.192503},
  url = {https://link.aps.org/doi/10.1103/PhysRevLett.125.192503}
}

@article{Li2024,
  title = {Unbound $^{28}${{O}}, the Heaviest Oxygen Isotope Observed: A Cutting-Edge Probe for Testing Nuclear Models},
  author = {Li, Jian Guo and Hu, Bai Shan and Zhang, Shuang and Xu, Fu Rong},
  year = 2024,
  month = feb,
  journal = {Nucl. Sci. Tech.},
  volume = {35},
  number = {2},
  pages = {21},
  issn = {2210-3147},
  doi = {10.1007/s41365-024-01373-w}
}

@article{Li2025dynamicTES,
  title = {Gamow shell model for dynamic Thomas-Ehrman shift in $^{16}\mathrm{Ne}/^{16}\mathrm{C}$ and $^{18}\mathrm{Mg}/^{18}\mathrm{C}$},
  author = {Li, K. H. and Li, J. G. and Michel, N. and Li, H. H. and Chen, N. and Ma, C. W. and Zuo, W.},
  journal = {Phys. Rev. C},
  volume = {111},
  issue = {1},
  pages = {014302},
  numpages = {8},
  year = {2025},
  month = {Jan},
  publisher = {American Physical Society},
  doi = {10.1103/PhysRevC.111.014302},
  url = {https://link.aps.org/doi/10.1103/PhysRevC.111.014302}
}

@misc{Lin2025,
  title = {Radii of Proton Emitters},
  author = {Lin, Y.R. and Wang, S.M. and Nazarewicz, W.},
  year = {2025},
  month = {nov},
  eprint = {2511.10238},
  archiveprefix = {arXiv},
}

@article{Ma2025,
  title = {Multi-Proton Emission at the Limits of Nuclear Stability: Challenges for Extreme Open Quantum Systems},
  author = {Ma, Yu Gang},
  year = 2025,
  month = oct,
  journal = {Nucl. Sci. Tech.},
  volume = {36},
  number = {12},
  pages = {236},
  issn = {2210-3147},
  doi = {10.1007/s41365-025-01831-z}
}

@article{Magilligan2020,
  title = {New isospin-breaking ``USD'' Hamiltonians for the sd shell},
  author = {Magilligan, A. and Brown, B. A.},
  journal = {Phys. Rev. C},
  volume = {101},
  issue = {6},
  pages = {064312},
  numpages = {15},
  year = {2020},
  month = {Jun},
  publisher = {American Physical Society},
  doi = {10.1103/PhysRevC.101.064312},
  url = {https://link.aps.org/doi/10.1103/PhysRevC.101.064312}
}

@article{Michel2004,
  title = {Proton-neutron coupling in the Gamow shell model: The lithium chain},
  author = {Michel, N. and Nazarewicz, W. and P\l{}oszajczak, M.},
  journal = {Phys. Rev. C},
  volume = {70},
  issue = {6},
  pages = {064313},
  numpages = {11},
  year = {2004},
  month = {Dec},
  publisher = {American Physical Society},
  doi = {10.1103/PhysRevC.70.064313},
  url = {https://link.aps.org/doi/10.1103/PhysRevC.70.064313}
}

@article{Michel2009,
author = {N. Michel and W. Nazarewicz and M. P{\l}oszajczak and T. Vertse},
title = {Shell model in the complex energy plane},
journal = {J. Phys. G},
volume = {36},
pages = {013101},
year = {2009},
doi = {10.1088/0954-3899/36/1/013101}
}

@book{Michel2021,
  author =        {Nicolas Michel and Marek P{\l}oszajczak},
  publisher =     {Springer, Cham},
  series =        {Lecture Notes in Physics},
  title =         {Gamow Shell Model, The Unified Theory of Nuclear
                   Structure and Reactions},
  volume =        {983},
  year =          {2021},
  doi =           {https://doi.org/10.1007/978-3-030-69356-5},
}

@article{Michel2021a,
  title = {Proton decays in $^{16}\mathrm{Ne}$ and $^{18}\mathrm{Mg}$ and isospin-symmetry breaking in carbon isotopes and isotones},
  author = {Michel, N. and Li, J. G. and Xu, F. R. and Zuo, W.},
  journal = {Phys. Rev. C},
  volume = {103},
  issue = {4},
  pages = {044319},
  numpages = {9},
  year = {2021},
  month = {Apr},
  publisher = {American Physical Society},
  doi = {10.1103/PhysRevC.103.044319},
  url = {https://link.aps.org/doi/10.1103/PhysRevC.103.044319}
}

@article{Michel2023,
  title = {Description of the Proton-Decaying ${0}_{2}^{+}$ Resonance of the $\ensuremath{\alpha}$ Particle},
  author = {Michel, N. and Nazarewicz, W. and P\l{}oszajczak, M.},
  journal = {Phys. Rev. Lett.},
  volume = {131},
  issue = {24},
  pages = {242502},
  numpages = {6},
  year = {2023},
  month = {Dec},
  publisher = {American Physical Society},
  doi = {10.1103/PhysRevLett.131.242502},
  url = {https://link.aps.org/doi/10.1103/PhysRevLett.131.242502}
}

@article{Miller2019,
title = {Proton superfluidity and charge radii in proton-rich calcium isotopes},
author = {Miller, A. J. and Minamisono, K. and Klose, A. and Garand, D. and Kujawa, C. and Lantis, J. D. and Liu, Y. and Maa{\ss}, B. and Mantica, P. F. and Nazarewicz, W. and N{\"o}rtersh{\"a}user, W. and Pineda, S. V. and Reinhard, P.-G. and Rossi, D. M. and Sommer, F. and Sumithrarachchi, C. and Teigelh{\"o}fer, A. and Watkins, J.},
volume = {15},
copyright = {20.900},
url = {https://www.nature.com/articles/s41567-019-0416-9},
doi = {10.1038/s41567-019-0416-9},
number = {5},
journal = {Nat. Phys.},
year = {2019},
pages = {432--436}
}

@article{Miyagi2022,
  title = {{Converged {\it ab initio} calculations of heavy nuclei}},
  author = {Miyagi, T. and Stroberg, S. R. and Navr\'atil, P. and Hebeler, K. and Holt, J. D.},
  journal = {Phys. Rev. C},
  volume = {105},
  issue = {1},
  pages = {014302},
  numpages = {14},
  year = {2022},
  month = {Jan},
  publisher = {American Physical Society},
  doi = {10.1103/PhysRevC.105.014302},
  url = {https://link.aps.org/doi/10.1103/PhysRevC.105.014302}
}

@article{Nazarewicz1996,
  title = {Structure of proton drip-line nuclei around doubly magic $^{48}\mathrm{Ni}$},
  author = {Nazarewicz, W. and Dobaczewski, J. and Werner, T. R. and Maruhn, J. A. and Reinhard, P.-G. and Rutz, K. and Chinn, C. R. and Umar, A. S. and Strayer, M. R.},
  journal = {Phys. Rev. C},
  volume = {53},
  issue = {2},
  pages = {740--751},
  numpages = {0},
  year = {1996},
  month = {Feb},
  publisher = {American Physical Society},
  doi = {10.1103/PhysRevC.53.740}
}

@article{Nazarewicz2025,
  title = {The lessons learned from ephemeral nuclei},
  author = {Nazarewicz, W. and Sobotka, L.},
  journal = {Phys.  Today},
  volume = {78},
  pages = {30},
  year = {2025},
  doi = {10.1063/pt.yvjv.skzx},
  url = {https://physicstoday.aip.org/features/the-lessons-learned-from-ephemeral-nuclei}
}

@article{Pfutzner2023,
  author={M. Pf\"utzner and I. Mukha and S. M. Wang},
  title={Two-proton emission and related phenomena},
  journal={Prog. Part. Nucl. Phys.},
  doi = {10.1016/j.ppnp.2023.104050},
  volume={123},
  pages={104050},
  year={2023}
}

@article{Phillips2025,
  title = {Clarifying the ${N},{Z}=14$ Shells near the Drip Lines from the Spectroscopy of $^{22}\mathrm{Si}$ and $^{21}\mathrm{Al}$},
  author = {Phillips, J. S. and Charity, R. J. and Dronchi, N. and Webb, H. and Sobotka, L. G. and Basson, M. J. and Benetti, C. and Brown, B. A. and Brown, K. W. and Brown, S. and Chung-Jung, J. and Cory, J. R. and Flores, G. and Gade, A. and Gajdosik, M. and Gillespie, S. and Kuich, M. and McCormick, C. E. and Parry, T. and Pereira, J. and Weisshaar, D. and Zerbach, V.},
  journal = {Phys. Rev. Lett.},
  volume = {135},
  issue = {25},
  pages = {252501},
  numpages = {8},
  year = {2025},
  month = {Dec},
  publisher = {American Physical Society},
  doi = {10.1103/v57q-45qj},
  url = {https://link.aps.org/doi/10.1103/v57q-45qj}
}

@article{Robert2012,
  title = {Medium-Mass Nuclei with Normal-Ordered Chiral $NN\mathbf{+}3N$ Interactions},
  author = {Roth, Robert and Binder, Sven and Vobig, Klaus and Calci, Angelo and Langhammer, Joachim and Navr\'atil, Petr},
  journal = {Phys. Rev. Lett.},
  volume = {109},
  issue = {5},
  pages = {052501},
  numpages = {5},
  year = {2012},
  month = {Jul},
  publisher = {American Physical Society},
  doi = {10.1103/PhysRevLett.109.052501},
  url = {https://link.aps.org/doi/10.1103/PhysRevLett.109.052501}
}

@article{Stroberg2021,
  title = {Ab Initio Limits of Atomic Nuclei},
  author = {Stroberg, S. R. and Holt, J. D. and Schwenk, A. and Simonis, J.},
  journal = {Phys. Rev. Lett.},
  volume = {126},
  issue = {2},
  pages = {022501},
  numpages = {6},
  year = {2021},
  month = {Jan},
  publisher = {American Physical Society},
  doi = {10.1103/PhysRevLett.126.022501},
  url = {https://link.aps.org/doi/10.1103/PhysRevLett.126.022501}
}

@article{Stanoiu2004,
  title = {$N=14$ and 16 shell gaps in neutron-rich oxygen isotopes},
  author = {Stanoiu, M. and Azaiez, F. and Dombr\'adi, Zs. and Sorlin, O. and Brown, B. A. and Belleguic, M. and Sohler, D. and Saint Laurent, M. G. and Lopez-Jimenez, M. J. and Penionzhkevich, Y. E. and Sletten, G. and Achouri, N. L. and Ang\'elique, J. C. and Becker, F. and Borcea, C. and Bourgeois, C. and Bracco, A. and Daugas, J. M. and Dlouh\'y, Z. and Donzaud, C. and Duprat, J. and F\"ul\"op, Zs. and Guillemaud-Mueller, D. and Gr\'evy, S. and Ibrahim, F. and Kerek, A. and Krasznahorkay, A. and Lewitowicz, M. and Leenhardt, S. and Lukyanov, S. and Mayet, P. and Mandal, S. and van der Marel, H. and Mittig, W. and Mr\'azek, J. and Negoita, F. and De Oliveira-Santos, F. and Podoly\'ak, Zs. and Pougheon, F. and Porquet, M. G. and Roussel-Chomaz, P. and Savajols, H. and Sobolev, Y. and Stodel, C. and Tim\'ar, J. and Yamamoto, A.},
  journal = {Phys. Rev. C},
  volume = {69},
  issue = {3},
  pages = {034312},
  numpages = {10},
  year = {2004},
  month = {Mar},
  publisher = {American Physical Society},
  doi = {10.1103/PhysRevC.69.034312},
  url = {https://link.aps.org/doi/10.1103/PhysRevC.69.034312}
}

@article{Stefan2014,
  title = {Probing nuclear forces beyond the drip-line using the mirror nuclei $^{16}\mathrm{N}$ and $^{16}\mathrm{F}$},
  author = {Stefan, I. and de Oliveira Santos, F. and Sorlin, O. and Davinson, T. and Lewitowicz, M. and Dumitru, G. and Ang\'elique, J. C. and Ang\'elique, M. and Berthoumieux, E. and Borcea, C. and Borcea, R. and Buta, A. and Daugas, J. M. and de Grancey, F. and Fadil, M. and Gr\'evy, S. and Kiener, J. and Lefebvre-Schuhl, A. and Lenhardt, M. and Mrazek, J. and Negoita, F. and Pantelica, D. and Pellegriti, M. G. and Perrot, L. and Ploszajczak, M. and Roig, O. and Saint Laurent, M. G. and Ray, I. and Stanoiu, M. and Stodel, C. and Tatischeff, V. and Thomas, J. C.},
  journal = {Phys. Rev. C},
  volume = {90},
  issue = {1},
  pages = {014307},
  numpages = {5},
  year = {2014},
  month = {Jul},
  publisher = {American Physical Society},
  doi = {10.1103/PhysRevC.90.014307},
  url = {https://link.aps.org/doi/10.1103/PhysRevC.90.014307}
}

@article{Sun2017,
title = {Resonance and continuum Gamow shell model with realistic nuclear forces},
author = {Z. H. Sun and Q. Wu and Z. H. Zhao and B. S. Hu and S. J. Dai and F. R. Xu},
journal = {Phys. Lett. B},
volume = {769},
pages = {227-232},
year = {2017},
issn = {0370-2693},
doi = {https://doi.org/10.1016/j.physletb.2017.03.054},
url = {https://www.sciencedirect.com/science/article/pii/S0370269317302459}
}

@article{Thomas1952,
  title = {An Analysis of the Energy Levels of the Mirror Nuclei, $^{}\mathrm{C}^{13}$ and $^{}\mathrm{N}^{13}$},
  author = {Thomas, R. G.},
  journal = {Phys. Rev.},
  volume = {88},
  issue = {5},
  pages = {1109--1125},
  numpages = {0},
  year = {1952},
  month = {Dec},
  publisher = {American Physical Society},
  doi = {10.1103/PhysRev.88.1109}
}

@article{Towner1987,
title = {Quenching of spin matrix elements in nuclei},
journal = {Phys. Rep.},
volume = {155},
number = {5},
pages = {263-377},
year = {1987},
issn = {0370-1573},
doi = {https://doi.org/10.1016/0370-1573(87)90138-4},
url = {https://www.sciencedirect.com/science/article/pii/0370157387901384},
author = {I.S. Towner},
}

@article{Volya2005,
  title = {Discrete and Continuum Spectra in the Unified Shell Model Approach},
  author = {Volya, Alexander and Zelevinsky, Vladimir},
  journal = {Phys. Rev. Lett.},
  volume = {94},
  issue = {5},
  pages = {052501},
  numpages = {4},
  year = {2005},
  month = {Feb},
  publisher = {American Physical Society},
  doi = {10.1103/PhysRevLett.94.052501},
  url = {https://link.aps.org/doi/10.1103/PhysRevLett.94.052501}
}

@article{Wang2019,
  title = {Structure and decay of the extremely proton-rich nuclei $^{11,12}\mathrm{O}$},
  author = {Wang, S. M. and Nazarewicz, W. and Charity, R. J. and Sobotka, L. G.},
  journal = {Phys. Rev. C},
  volume = {99},
  issue = {5},
  pages = {054302},
  numpages = {8},
  year = {2019},
  month = {May},
  publisher = {American Physical Society},
  doi = {10.1103/PhysRevC.99.054302}
}

@article{Wang2021,
  title = {Fermion Pair Dynamics in Open Quantum Systems},
  author = {Wang, S. M. and Nazarewicz, W.},
  journal = {Phys. Rev. Lett.},
  volume = {126},
  issue = {14},
  pages = {142501},
  numpages = {6},
  year = {2021},
  month = {Apr},
  publisher = {American Physical Society},
  doi = {10.1103/PhysRevLett.126.142501},
  url = {https://link.aps.org/doi/10.1103/PhysRevLett.126.142501}
}

@article{Wang2025,
    author = "Wang, J. L. and Xie, M. R. and Li, K. H. and Wang, P. Y. and Michel, N. and Yuan, Q. and Li, J. G.",
    title = "{Gamow shell model predictions for six-proton unbound nucleus 20Si}",
    eprint = "2512.07594",
    archivePrefix = "arXiv",
    primaryClass = "nucl-th",
    doi = "10.1016/j.physletb.2025.140030",
    journal = "Phys. Lett. B",
    volume = "872",
    pages = "140030",
    year = "2026"
}

@article{AME2021,
	author = {Meng Wang and W. J. Huang and F. G. Kondev and G. Audi and S. Naimi},
	doi = {10.1088/1674-1137/abddaf},
	journal = {Chin. Phys. C},
	month = {mar},
	number = {3},
	pages = {030003},
	publisher = {Chinese Physical Society and the Institute of High Energy Physics of the Chinese Academy of Sciences and the Institute of Modern Physics of the Chinese Academy of Sciences and IOP Publishing Ltd},
	title = {The {AME} 2020 atomic mass evaluation ({II}). {T}ables, graphs and references},
	url = {https://dx.doi.org/10.1088/1674-1137/abddaf},
	volume = {45},
	year = {2021}
}

@article{Webb2019,
  title = {First Observation of Unbound $^{11}\mathrm{O}$, the Mirror of the Halo Nucleus $^{11}\mathrm{Li}$},
  author = {Webb, T. B. and Wang, S. M. and Brown, K. W. and Charity, R. J. and Elson, J. M. and Barney, J. and Cerizza, G. and Chajecki, Z. and Estee, J. and Hoff, D. E. M. and Kuvin, S. A. and Lynch, W. G. and Manfredi, J. and McNeel, D. and Morfouace, P. and Nazarewicz, W. and Pruitt, C. D. and Santamaria, C. and Smith, J. and Sobotka, L. G. and Sweany, S. and Tsang, C. Y. and Tsang, M. B. and Wuosmaa, A. H. and Zhang, Y. and Zhu, K.},
  journal = {Phys. Rev. Lett.},
  volume = {122},
  issue = {12},
  pages = {122501},
  numpages = {6},
  year = {2019},
  month = {Mar},
  publisher = {American Physical Society},
  doi = {10.1103/PhysRevLett.122.122501}
}

@article{Wu2021,
  title = {$\ensuremath{\beta}$-decay spectroscopy of the proton drip-line nucleus $^{22}\mathrm{Al}$},
  author = {Wu, C. G. and Wu, H. Y. and Li, J. G. and Luo, D. W. and Li, Z. H. and Hua, H. and Xu, X. X. and Lin, C. J. and Lee, J. and Sun, L. J. and Liang, P. F. and Yuan, C. X. and Yang, Y. Y. and Wang, J. S. and Wang, D. X. and Duan, F. F. and Lam, Y. H. and Ma, P. and Gao, Z. H. and Hu, Q. and Bai, Z. and Ma, J. B. and Wang, J. G. and Zhong, F. P. and Jiang, Y. and Liu, Y. and Hou, D. S. and Li, R. and Ma, N. R. and Ma, W. H. and Shi, G. Z. and Yu, G. M. and Patel, D. and Jin, S. Y. and Wang, Y. F. and Yu, Y. C. and Zhou, Q. W. and Wang, P. and Hu, L. Y. and Fan, S. Q. and Wang, X. and Zang, H. L. and Li, P. J. and Zhao, Q. Q. and Yang, L. and Wen, P. W. and Yang, F. and Jia, H. M. and Zhang, G. L. and Pan, M. and Wang, X. Y. and Sun, H. H. and Hu, Z. G. and Liu, M. L. and Chen, R. F. and Yang, W. Q. and Hou, S. Q. and He, J. J. and Zhao, Y. M. and Xu, F. R. and Zhang, H. Q.},
  collaboration = {RIBLL Collaboration},
  journal = {Phys. Rev. C},
  volume = {104},
  issue = {4},
  pages = {044311},
  numpages = {9},
  year = {2021},
  month = {Oct},
  publisher = {American Physical Society},
  doi = {10.1103/PhysRevC.104.044311},
  url = {https://link.aps.org/doi/10.1103/PhysRevC.104.044311}
}

@article{Wylie2021,
  title = {Spectroscopic factors in dripline nuclei},
  author = {Wylie, J. and Oko\l{}owicz, J. and Nazarewicz, W. and P\l{}oszajczak, M. and Wang, S. M. and Mao, X. and Michel, N.},
  journal = {Phys. Rev. C},
  volume = {104},
  issue = {6},
  pages = {L061301},
  numpages = {7},
  year = {2021},
  month = {Dec},
  publisher = {American Physical Society},
  doi = {10.1103/PhysRevC.104.L061301},
  url = {https://link.aps.org/doi/10.1103/PhysRevC.104.L061301}
}

@article{Xing2025,
  title = {${Z}=14$ Magicity Revealed by the Mass of the Proton Dripline Nucleus $^{22}\mathrm{Si}$},
  author = {Xing, Y. M. and Luo, Y. F. and Zhang, Y. H. and Wang, M. and Zhou, X. H. and Li, J. G. and Li, K. H. and Yuan, Q. and Niu, Y. F. and Guo, J. Y. and Pei, J. C. and Xu, F. R. and de Angelis, G. and Litvinov, Yu. A. and Blaum, K. and Tanihata, I. and Yamaguchi, T. and Yu, Y. and Zhou, X. and Xu, H. S. and Chen, Z. Y. and Chen, R. J. and Deng, H. Y. and Fu, C. Y. and Ge, W. W. and Huang, W. J. and Jiao, H. Y. and Li, H. F. and Liao, T. and Shi, J. Y. and Si, M. and Sun, M. Z. and Shuai, P. and Tu, X. L. and Wang, Q. and Xu, X. and Yan, X. L. and Yuan, Y. J. and Zhang, M.},
  journal = {Phys. Rev. Lett.},
  volume = {135},
  issue = {1},
  pages = {012501},
  numpages = {8},
  year = {2025},
  month = {Jul},
  publisher = {American Physical Society},
  doi = {10.1103/ffwt-n7yc},
  url = {https://link.aps.org/doi/10.1103/ffwt-n7yc}
}

@article{Zhang2022,
title = {The roles of three-nucleon force and continuum coupling in mirror symmetry breaking of oxygen mass region},
journal = {Phys. Lett. B},
volume = {827},
pages = {136958},
year = {2022},
issn = {0370-2693},
doi = {https://doi.org/10.1016/j.physletb.2022.136958},
url = {https://www.sciencedirect.com/science/article/pii/S0370269322000922},
author = {S. Zhang and Y. Z. Ma and J. G. Li and B. S. Hu and Q. Yuan and Z. H. Cheng and F. R. Xu},
}
\end{document}